\documentclass[preprint,...]{revtex4-1}

\usepackage{graphicx}

\usepackage{amsmath}
\usepackage{amsfonts}
\usepackage{amssymb}
\usepackage{calrsfs}

\usepackage{appendix}
\usepackage{emptypage}
\usepackage{cancel}
\usepackage{fancyvrb}
\usepackage{gensymb}
\usepackage{verbatim}
\usepackage{wrapfig}
\usepackage{units}
\usepackage{bm}
\usepackage{url}
\usepackage{xfrac}
\usepackage{xcolor}
\usepackage[english]{babel}
\usepackage{blindtext}
\usepackage{setspace}
\usepackage{csquotes}
\usepackage{xcolor}
\usepackage{soul}

\usepackage{array}
\newcolumntype{L}[1]{>{\raggedright\let\newline\\\arraybackslash\hspace{0pt}}m{#1}}
\newcolumntype{C}[1]{>{\centering\let\newline\\\arraybackslash\hspace{0pt}}m{#1}}
\newcolumntype{R}[1]{>{\raggedleft\let\newline\\\arraybackslash\hspace{0pt}}m{#1}}

\let\baraccent=\= 
\renewcommand{\=}[1]{\stackrel{#1}{=}} 

  \newcommand{\lp}{\left(}
  \newcommand{\rp}{\right)}

  \newcommand{\fpsi}{f \lp \psi \rp}
  \newcommand{\hyperp}{\theta_{hp}}
  
  \newcommand{\data}{\vec d}
  
  \let\given\givenbase

  \newcommand{\cpposterior}{\lp \fpsi \given \data \rp}

  \newcommand{\cplikelihood}{\lp \data \given \fpsi \rp}

  \newcommand{\cpprior}{\lp \fpsi \rp}

\begin{document}
	

\title{Single Gaussian Process Method for Arbitrary Tokamak Regimes with a Statistical Analysis}


\author{J.Leddy$^{1}$}
\email[]{jleddy@txcorp.com}
\author{S.Madireddy$^2$}
\author{E.Howell$^{1}$}
\author{S.Kruger$^{1}$}
\affiliation{$^1$Tech-X Corporation, Boulder CO USA}
\affiliation{$^2$Argonne National Laboratory}

\date{\today}

\begin{abstract}
	Gaussian Process Regression (GPR) is a Bayesian method for inferring profiles based on input data. The technique is increasing in popularity in the fusion community due to its many advantages over traditional fitting techniques including intrinsic uncertainty quantification and robustness to over-fitting. This work investigates the use of a new method, the change-point method, for handling the varying length scales found in different tokamak regimes.  The use of the Student's t-distribution for the Bayesian likelihood probability is also investigated and shown to be advantageous in providing good fits in profiles with many outliers.  To compare different methods, synthetic data generated from analytic profiles is used to create a database enabling a quantitative statistical comparison of which methods perform the best.  Using a full Bayesian approach with the change-point method, Mat\'ern kernel for the prior probability, and Student's t-distribution for the likelihood is shown to give the best results.
\end{abstract}


\maketitle

\section{Introduction} \label{sec:intro}
Achieving fusion energy using magnetic confinement requires understanding plasmas that are hotter than the sun and their relationship to the magnetic field.  This is not an easy task because diagnostics in this harsh environment requires understanding the physics of the diagnostic, and the plasma itself.  This is known as an inverse problem:  understanding the plasma requires understanding the diagnostic, which requires understanding the plasma.
	
For magnetic confinement, the small Larmor radius expansion gives to first order force balance: $J \times B=\nabla p $.~\cite{HazeltineMeiss}  In tokamaks, this gives the Grad-Shafranov Equation (GSE) and solving the inverse problem described above is known as equilibrium reconstruction.~\cite{lao85,lao05}  Equilibrium reconstruction has emerged as a critical tool for tokamak control~\cite{ferron1998real} and in  plasma performance optimization~\cite{Strait_1994,Levinton_1995,Strait_1995,Zwingmann_2003}. The most widely-used code for equilibrium reconstruction is EFIT~\cite{Lao_1985,Lao_2005}, which has been used at the DIII-D~\cite{Lao_1990,Lao_2005}, C-MOD~\cite{In_2000}, JET~\cite{OBrien_1992}, HIT~\cite{Nelson_1994}, START~\cite{Sykes_1999,Appel_2001}, MAST~\cite{Appel_2006}, KSTAR~\cite{Lee_1999,Park_2011}, JT-60U~\cite{Oikawa_2000}, NSTX~\cite{Sabbagh_2001}, TORE SUPRA~\cite{Zwingmann_2003,Li_2011}, HT-7~\cite{Li_2003}, HL-2A~\cite{Hongda_2006}, and EAST~\cite{Jinping_2009,Li_2013} devices.   Equilibrium reconstruction is a critical part of the development of the fusion.
	
However, while traditional equilibrium reconstructions using only the magnetic data can be robust and routine, using additional data such as Motional Stark Effect (MSE)~\cite{Levinton_1995}, Thomson scattering diagnostic~\cite{Carlstrom92,Carlstrom_1995}, or soft X-Ray~\cite{Christiansen_1989} are more problematic both due to the additional complexity additional of modeling internal diagnostics and diagnostic limitations.  The best equilibrium reconstructions typically require considerable manual intervention.   For burning plasmas, where the harsh requirements and increased need for plasma control, more robust methods are needed.
	
Using a Bayesian approach to these problems offers many advantages~\cite{fischer2004,fischer2004b}.  It more naturally allows one to account for the uncertainty in the data and for prior knowledge to be incorporated into the problem.  The specific technique used in this paper will be Gaussian Process Regression, a technique that is becoming widely used in the uncertainty quantification and machine learning~\cite{williams2006gaussian}.  It's introduction into the fusion community was by Svensson~\cite{svensson2011non} for soft X-ray tomography, but has since grown to include other diagnostics~\cite{svensson2011non,li2016,joung2018,kwak2020bayesian,kwak2021bayesian}.   It's use within the fusion community has broadened to be used for verification and validation~\cite{chilenski2015,fischer2020,ho2019,mathews2021}, and in equilibrium reconstruction as well~\cite{Howell_2020,kwak2021equilibria}.  The latter is our ultimate goal.
	
One of the biggest drawbacks in using GPR is the same as that using traditional fitting methods:  how to account for very different profiles in the fitting.   Most notably for tokamak fitting is the difference between the L-mode (low-confinement mode) and H-mode regimes.  The H-mode regime is characterized by an edge pedestal that has a transport barrier giving short radial scale lengths.  In the terminology of GPR (discussed more below), there is a {\em non-stationary} hyperparameters.  That is, the short radial scale lengths imply that the radial correlation of the observations change.   There are a few ways of handling this.  The most popular method both within the larger ML community is known as {\em warping}, and is equivalent to a coordinate transformation in applied math, or non-uniform grids in numerical analysis.  In the fusion community, a  {\em tanh} warping function was used to develop non-stationary hyperparameters for H-mode profiles was introduced by Chilenski~\cite{chilenski2015}.  This has been used by many authors~\cite{li2016,kwak2020bayesian,sciortino2020,fischer2020,ho2019,kwak2021bayesian} since then. 


The disadvantage of the warping approach to radially varying hyperparameters is the same problem that confronts the traditional fitting approach:  one needs to know the type of fitting to perform ahead of time.  This problem has been addressed in multiple ways.  One method is to determine if the profile is L-mode or H-mode ahead of time, and then perform the appropriate traditional fitting~\cite{xing2021}, or hyperparameter fitting~\cite{mathews2021}.  In this paper, we explore a different approach using a change-point kernel with the goal of using a single method regardless of the tokamak parameter regime.  Our methodology for these investigations will also be different:  we will focus strictly on synthetic analytic profiles using synthetically-generated data.  Using this method allows us to analytically explore the effects of the distribution of the noise and the influence of outliers in the data in a quantitative way.
	
Our paper proceeds as follows.  We first briefly review the basics of Gaussian Process Regression to introduce the basic concepts and terminology.  Although this has been discussed in the fusion literature above, as well as other fields such as machine learning~\cite{williams2006gaussian}, we will be introducing new methods and need to clarify the context of these new methods.  We next apply our new methods to the synthetic data characteristic of tokamak regimes and contrast with prior methods in the fusion community.  Finally, we draw conclusions and point to future work.

\section{Background}
\label{sec:background}
Traditional method to \textit{fitting} a profile $\fpsi$ typically involves
finding a the function that depends on a set of \textit{parameters} and minimizing the error.  For example, one can write $\fpsi = \sum f_i \alpha_i$, where $f_i$ are the
parameters, and $\alpha_i$ is the \textit{fitting function}, and then minimize the error
related to the data.  This is the traditional method of EFIT which uses spline functions for the fitting functions with L-mode profiles, and splines and a \emph{tanh} function for H-mode profiles.  Advanced techniques for fitting using the OMFIT~\cite{OMFIT2015} are discussed in Ref.~\cite{Logan2018}\footnote{This paper discusses the OMFIT profiles module that can also call out to the \texttt{gptools} software package that forms the basis of the paper by Chilenski~\cite{chilenski2015}.}

In the Bayesian view, the process is fundamentally different.  One first
calculates the probability of a function being correct, and the most probable
function is used as that fit.   Mathematically, this is written as the
$p\cpposterior$:  the probability of $\fpsi$ given the data $\data$.  This is a
conceptual shift compared to traditional fitting.  It acknowledges that there is
uncertainty in the data, and uncertainty in $\fpsi$ itself.   To find this
probability is not straightforward directly.  Instead, we use Bayes' Theorem:
\begin{equation}
  \label{BayesTheorem}
  \underbrace{p\cpposterior}_{\text{posterior}} = 
      \frac{\overbrace{p\cplikelihood}^{\text{likelihood}}}
            {\underbrace{p\lp \data \rp}_{\text{marginal likelihood}}}
      \underbrace{p\lp \fpsi \rp}_{\text{prior}}
\end{equation}
The advantage of Bayes' theorem in practical terms is that it easier to compute
the probability of the likelihood and prior distribution than the posterior
likelihood.  The marginal likelihood is a normalizing constant
that ensures the total probability is one.

The first step that we wish to consider is how to encode the space of all
functions into the prior probability.   Above, we discussed how parameters were used to
parameterize a function.  A similar method is used to explore the distribution
space of the probability functions here.  In PDE's, exploring a functional space
is done using discretizations (e.g., finite difference).  The problem with this
for probability functions is that although we might wish to assume that $\fpsi$
lies within a functional space (e.g., Hilbert space), the data, which is subject
to noise, does not.  Using traditional PDE discretizations would also subject
one to systematic bias in determining the distributions.  Hence, randomized
sampling of the distribution space is generally preferred.  Two methods for
exploring this space are polynomial chaos expansions and stochastic processes.
In both methods, the conversion of the continuum form of the probabilities above
are converted to discrete form. For distributions, \textit{hyperparameters}, denoted $\hyperp$,are
used to explore the probability space for the prior and likelihood functions.

In this paper, we will consider a specific form of a stochastic process known as
the Gaussian Process.   Intuitively, a Gaussian process is one whose random
numbers generated (definition of a stochastic process) is distributed normally.
Mathematically, we write the prior sampling as
\begin{equation}
    \label{eq:priorGP}
    p\cpprior = GP\lp \mu\lp\psi\rp, K\lp \psi, \psi^\prime\rp \rp
\end{equation}
where $\mu$ is the mean function of the process and $K$ is the covariance function,
or kernel.  A common Gaussian process uses a zero mean and the  squared exponential kernel (a.k.a. radial basis function):
\begin{equation}
    \label{eq:Ksek}
	    K_{sek}\left(\psi,\psi'\right) = \theta_v e^ {-\frac{\left(\psi-\psi'\right)^2}{2\theta_l^2}}.
\end{equation}
The Gaussian process is distribution over all smooth functions of $\psi$. Each
function in the Gaussian process occurs with a certain probability that depends
on the two hyperparameters $\theta_v$ and $\theta_l$. The hyperparameter
$\theta_v$ characterizes how rapidly the probability decreases as the function departs from the
mean function ($0$ in this case), and $\theta_l$ characterizes the spatial
correlation of the function at two points $\psi$ and $\psi'$. A kernel is
\textit{stationary} if it only depends on the distance between two points
$K\left(\psi,\psi'\right)=K\left(\left|\psi-\psi'\right|\right)$, otherwise the
kernel is \textit{non-stationary}.   

In the context of fitting profiles of fusion data, the discussion of whether a
Gaussian process is stationary or not relates to the characteristic length scale
of the profile, or data, to be inferred.  For H-mode profiles, the short length
scale of the pedestal region motivates the use of non-stationary kernels;
e.g., as used in Chilenski~\cite{chilenski2015}.   Similarly, inferring magnetic
data can be challenging because of rapid changes in the data requiring
non-stationary kernels~\cite{joung2018}.

%

The hyperparameters play an important role in GPR, and the choice of
hyperparameters requires special consideration. The simplest approaches is to
pick reasonable values for the hyperparameters that are learned from prior
experiments. In this approach, a large enough data set can overcome a bad choice
of hyperparameters. However, this approach suffers if underlying profile
changes drastically in different experimental regimes or if the amount of
measured data is limited, both of which are case here.  Thus, this approach is
not used in practice.
	
A second approach is the \textit{Empirical Bayesian} where one uses the data to determine 
the optimal hyperparameters.  There is a family of methods in this approach. One method chooses the likelihood
to also be a Gaussian function.  The combination of Gaussian functions in both the likelihood,
and a squared-exponential kernel enables an analytically-tractable method for
maximizing the marginal likelihood.   This method was used in the first introduction of GPR to the.~\cite{svensson2011non}
When the likelihood is not a Gaussian, a closed-form
analytic solution in general is not possible, and one must use numerical methods for this.

A different philosophical approach is the \textit{Full Bayesian} approach in which the
prior distribution is fixed before any data is observed.  In this approach, the
prior and likelihood are still described in terms of hyperparameters as
described above, but the full hyperparameter space is explored numerically by integrating out the
hyperparameters.  Specifically, the likelihood and prior probability functions
in Eq.~\ref{BayesTheorem}
are parameterized in terms of hyperparameters, and then the hyperparameters are numerically integrated out in a process known as \textit{marginalization}.   There a wide variety of
approaches to marginalization or in using partial approximations to accelerate
this integration.   In this paper, we will use \textit{Markov chain Monte Carlo}
methods (MCMC) to systematically perform the Bayesian integration.  In the fusion community,
Chilenski is a notable paper in discussion the Full Bayesian technique as
well as using non-stationary hyperparameters.  Empirical Bayesian remains popular
however because of the speed of it as compared to Full Bayesian techniques.  This
is discussed in more detail later.

In this paper, we will systematically explore different approaches to
using Bayesian fitting of profiles of varying length scales, and using both
Empirical Bayes and Full Bayes using MCMC.   The use of Full Bayes will be
required as we explore the use of different likelihood functions in fitting
noisy data.  We turn to the method of determining the best method next.

\section{Overview of Methods}

\subsection{Generation of synthetic data for tokamak parameter regimes}
Our goal is to understand generic issues related to using the GPR method to infer the the density ($n$) and temperature ($T$) profiles of tokamaks from localized measurements such as Thomson scattering. However, the profiles characteristic of L-mode, H-mode, ITB regimes can have varying length scales providing challenges to the GPR method.  Experimentally, tokamaks account for this varying resolution by having TS systems with a spatial resolution that varies across the domain in order to capture the sharp features associated with a pedestal. For example, the DIII-D tokamak core TS system has a coarser spatial resolution which is adequate for resolving the slowly varying core $n$ and $T$ profiles and a finer resolution in pedestal to resolve the sharp profile gradients in this region\cite{Carlstrom92}.

Here, rather than use raw experimental data as in prior work~\cite{svensson2011non,li2016,joung2018,kwak2020bayesian,kwak2021bayesian}, we create synthetic data to enable a more accurate understanding of the errors associated with different GPR methods.  Specifically, for L-mode profiles we use
\begin{equation}
    \label{Lmode-profile}
               \fpsi = f_o \lp 1 - \psi^{\alpha_1}\rp^{\alpha_2} + f_{edge}
\end{equation}
where the input parameters are $f_o$ (value at the center), $f_{edge}$ (value
at edge), $\alpha_1$ (controls the gradient near the center, and $\alpha_2$
controls the gradient near the edge.  The H-mode profile is given by:
\begin{equation}
    \label{Hmode-profile}
    \fpsi = f_o \lp 1 - \psi^{\alpha_1}\rp^{\alpha_2} + f_{edge} 
            + \frac{1}{2} f_{ped} 
            \lp 1. - \tanh{\frac{\psi - \psi_{ped}}{w_{ped}}}\rp
\end{equation}
where $f_{ped}$ and $w_{ped}$ control the pedestal height and width.

The synthetic data is created by taking 40 points uniformly across the domain, and then another 40 points uniformly across the pedestal region to account for the way in which experimental measurements are often localized in the pedestal.  The data at the discrete profile points is then perturbed using a 
random numbers to generate noise assuming a Gaussian distribution with a specified width (standard deviation).  
In this paper, the width varies from 10\% to 33\% of the data value itself.  In addition, the noise can be shifted uniformly; i.e., given a non-zero mean in the Gaussian, to represent systematic errors.  Outliers are 
synthetically created by choosing random points and using random values that
have a Gaussian width of 2 - 3 times the data value. If a negative outlier is generated, then
it is made positive.  As seen below, this can create an additional systematic error.  The synthetic ITB profiles 
are similar to the H-mode profiles except that the \emph{tanh} profile is moved to 
the center of the plasma.

\subsection{Choosing a kernel} \label{sec:kernel}
As discussed in Section~\ref{sec:background}, the standard basic kernel for GPR is the squared exponential kernels as given by Eq.~\ref{eq:Ksek}.  The hyperparameters in these kernels do not have to be chosen \emph{a priori}, but can instead be inferred from the data themselves.  Typically, a stationary kernel is used, but for H-mode profiles, Chilenski~\cite{chilenski2015}, using a a non-stationary kernel derived by Gibbs~\cite{Gibbs1997}, used a tanh function for the hyperparameters.  This addition can make the computation of the fit expensive, and additionally must be constrained or convergence becomes difficult.  We desire a robust fitting routine, so we opted for a second option that is available in GPflow~\cite{buckingham2008comparative}, the software that we use, called the change-point kernel.  This is effectively a piecewise function for the kernel where multiple individual kernels can be defined and locations can be specified for when one kernel changes to another.  The derivatives of the total kernel is ensured to be continuous by the use of a transfer function between the piecewise parts of the kernel, which smoothly transitions between the sub-kernels.
	
\begin{figure}
		\begin{center}
			\includegraphics[width=\textwidth]{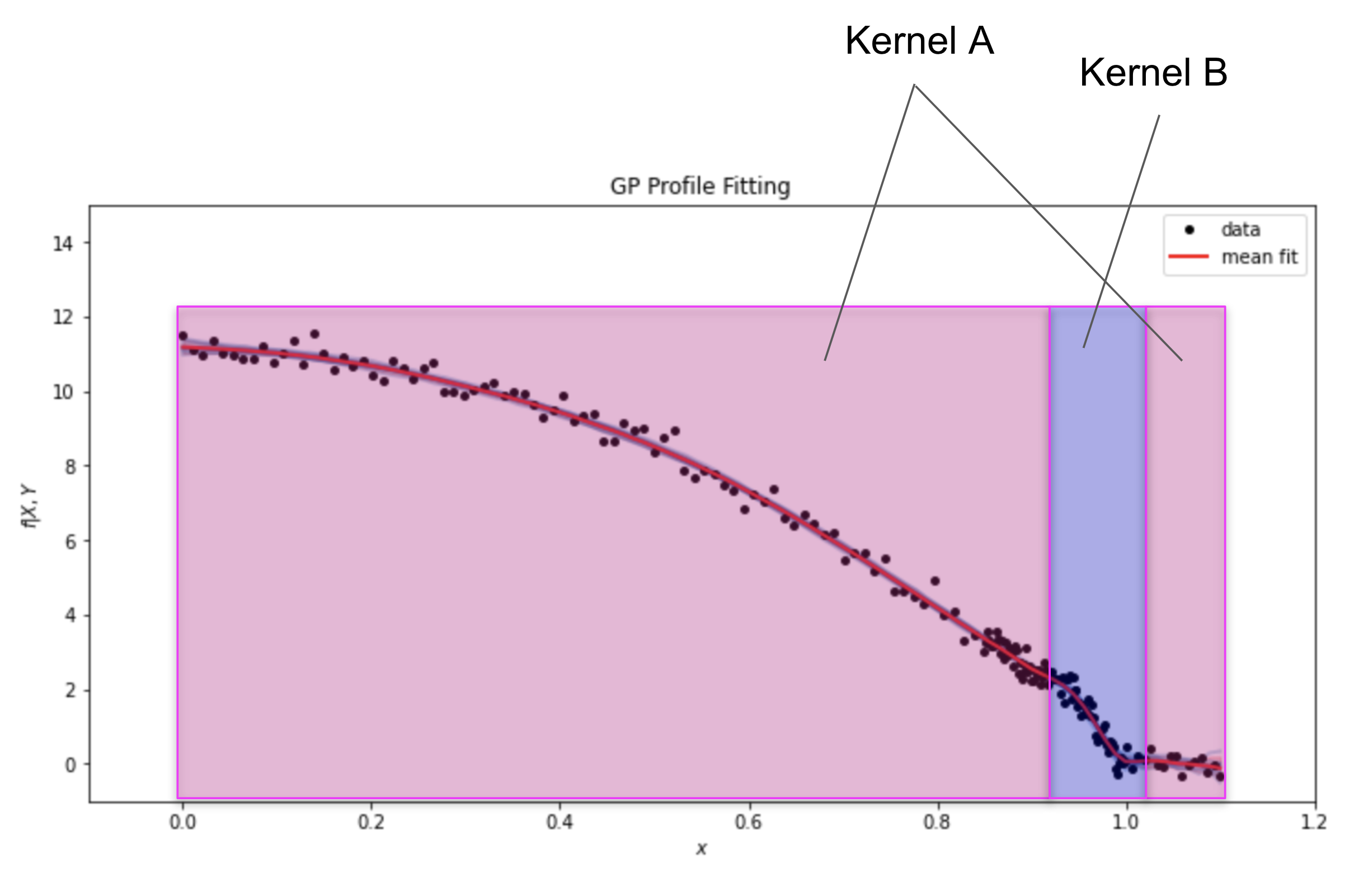}
		\end{center}
		\caption{An example profile of the synthetic data generated from that analytic profile are shown. The regions over which the parts of the change-point kernel are active are highlighted.  Kernel A and kernel B have an independent set of hyperparameters, and so can be tuned to fit the data appropriately.}
		\label{fig:cpkernel}
	\end{figure}

The H-mode pedestal occurs in the edge region and so we allow one kernel to describe $\psi \in [0,0.9]$ and $\psi \in [1.0,1.1]$, and then a separate kernel defines $\psi \in [0.9,1.0]$.  Each of these distributions has independent hyperparameters and therefore can represent a profile that has varying correlation length across the profile.  Because we use the data to infer the hyperparameter values, this actually also works well to describe an L-mode profile that has  a single effective correlation length.  In that case, both kernels would end up with similar values for the hyperparameters.  Figure \ref{fig:cpkernel} demonstrates the way this kernel has been constructed, as well as the data that illustrate the need for the change-point kernel.
	
The individual kernels chosen for kernels A and B (as indicated in figure \ref{fig:cpkernel}) are the well-established Mat\'ern covariance functions.  These are designed to indicate the correlation between points given some distance, $d=|\psi-\psi'|$, via the following equation:
	\begin{equation} \label{eq:Matern}
	    C_{5/2}(d) = \theta_v^2 \left( 1 + \frac{\sqrt{5}d}{\theta_l} + \frac{5d^2}{3\theta_l^2} \right) \exp\left( {-\frac{\sqrt{5}d}{\theta_l}} \right).
	\end{equation}
In this formula, $\theta_v$ is indicative of the width of the covariance function (like the standard deviation in a gaussian) and $\theta_l$ is a normalization parameter that indicates the degree of the correlation between points.  These two values are hyperparameters of the kernels that can be independently optimized for each kernel (A \& B).  In regions of large gradients/fluctuations we expect $\sigma_l$ to be small, while in regions of small gradients we expect it to be larger.

\subsection{Handling outliers} \label{sec:student}
Outlying data can negatively affect the fit and must be handled with care.  One obvious but labor intensive approach is to manually go through the data to identify outliers or ``bad" data and remove them.  This is, however, not ideal for multiple reasons.  Firstly, it takes time and effort, but we would like the fitting to be an automated process.  Secondly, sometimes data can look bad but be accurate, and it is not always possible to determine which is the case.  Finally, we would rather be able to incorporate all the data but keep track of the errors instead of eliminating data because they look bad.  Therefore, we conclude that we want to keep all the data, but we do not want outliers to negatively affect the fit.  Instead we want them to affect the error of the fit. In the Bayesian framework, this is handled by the likelihood probability.
	
In general, when letting the data inform the choice in hyperparameters, we want to maximize the log likelihood of the fit given the data.  For data with normally distributed noise, a squared exponential likelihood is used as it is appropriate, and it comes with the added benefit that the problem can be reduced analytically to an easily tractable form.  However, if there are outliers in the data, and/or the noise is non-Gaussian this is no longer an appropriate choice for the likelihood function.  A heavy-tailed distribution can capture the error on the outliers without affecting the quality of the fit for the data points with lower error.
	
	We have tested multiple heavy-tailed distributions including the Student's t-, logistic, and Laplace distributions.  All of these function adequately, as we will show, but the Student's t-distribution is the one performs best.  It is given by:
	\begin{equation}
		L(x) = \frac{\Gamma \left(\frac{\nu+1}{2} \right)}{\sqrt{\nu \pi}\Gamma \left(\frac{\nu}{2} \right)} \left(1 + \frac{x^2}{\nu} \right)^{-\frac{\nu+1}{2}}
	\end{equation}
	where $\nu$ is the degrees of freedom and $\Gamma$ is the gamma function.  This is a particularly nice function because it approaches a Gaussian as $\nu \rightarrow \infty$.  We can allow the degrees of freedom hyperparameter to be optimized during the GPR so that if the errors on the data are genuinely Gaussian then the value for $\nu$ will end up being large and this will capture the errors properly.  However, if we have outliers that would skew the fit, the $\nu$ parameter will adjust automatically to be smaller, increasing the weight of the likelihood function tails, allowing the quality of fit to remain high while minimizing over-fitting.  This does not come without a cost, however, since we can no longer make the same analytic simplifications to optimize hyperparameters that can be done with a pure Gaussian likelihood.  Instead, we resort to Markov Chain Monte Carlo (MCMC) to sample the hyperparameters space of both the kernel and the likelihood simultaneously.  Instead of optimizing, then, we obtain a distribution for each of the hyperparameters, and then sample from them to obtain possible fits.  The final fit we choose is the mean of all the sampled fits.
	
\section{Results}
We will discuss two regimes that we fit explicitly: L-mode and H-mode.  We do this because these two cases are illustrative of the reasoning behind our choice of kernel and likelihood functions.  Then we will go on to discuss the statistics of many fits.
	
\subsection{Fitting in various regimes}
	In all of the regimes we discuss below, the kernel and likelihood function are identical.  The automated tuning of the hyperparameters is all that differs, and so we will discuss those results and show the distributions for all the hyperparameters.  In this way, all of these regimes are fit without any human intervention or regime detection algorithms.  The data themselves inform the choice of hyperparameters for the GPR.
	
	\subsubsection{L-mode} \label{sec:lmode}
	
\begin{figure}
		\begin{center}
			\includegraphics[width=\textwidth]{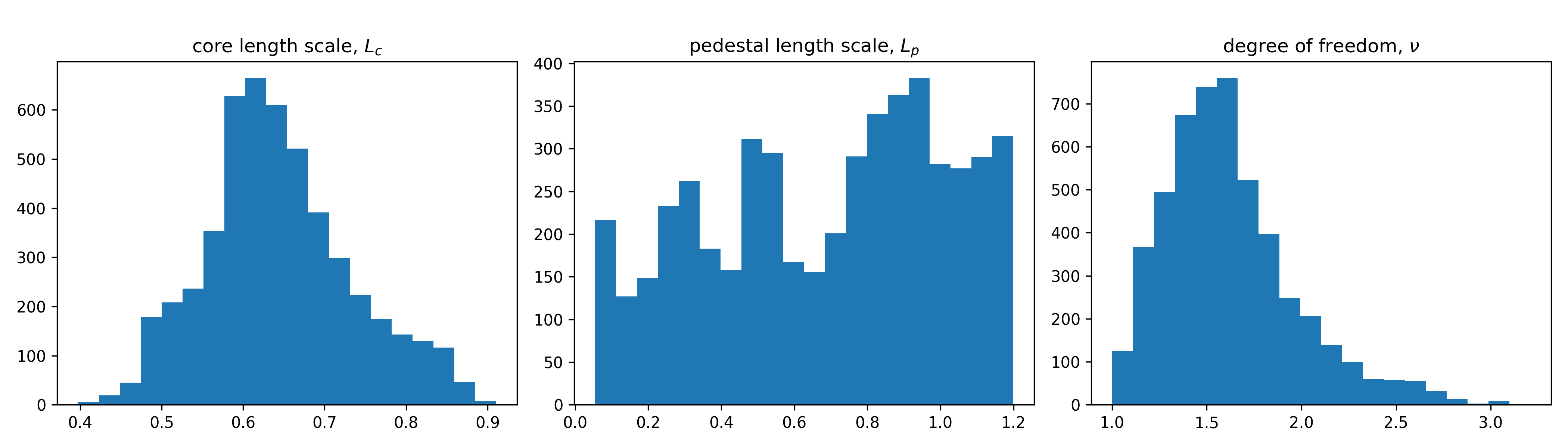}
		\end{center}
		\caption{Histograms of the hyperparameter distributions as found by the MCMC algorithm for the L-mode fit.  The y-axes are counts from the MCMC chain steps and the x-axes are the values for the variable in the title of the plots.  Although the analytic profile has a smoothly-varying length scale and fixed length scale, several length scales can be found in the noisy data illustrating the problems of standard data fitting techniques.}
		\label{fig:lmode-hp}
	\end{figure}

	In the L-mode, as discussed above, the profiles are smooth.  We can hypothesize the affect of such a profile on the hyperparameter distributions.  Firstly, it is expected that the hyperparameters of both kernels are similar since the correlation length across the profiles is roughly unchanging.  Additionally, we expect the degrees of freedom parameter to roughly correlate with the number and size of the outlying data.
	Figure \ref{fig:lmode-hp} shows the distributions for some of the hyperparameters as sampled using MCMC.  The hyperparameter $L_p$ represents the correlation length in what would be the pedestal region in H-mode ($\psi \in [0.9,1]$) and corresponds to the hyperparameter $\theta_l$ for the pedestal kernel in equation \ref{eq:Matern}.  This peaks around 0.8, but is roughly uniform across the 0.1-1.2 range.  A value of 0.1 is of similar size to the pedestal width itself indicating that all values over this result in a smooth fit across the pedestal region.  The hyperparameter $L_c$ represents the correlation length for the core (corresponding to the hyperparameter $\theta_l$ for the core kernel in equation \ref{eq:Matern}) and edge region and peaks around 0.5.  Due to the slight kink even in L-mode profiles, we do expect to see this small decrease in correlation length in the pedestal region, but for H-mode we would expect to see this value to be much lower to account for the steep pedestal.  The degrees of freedom hyperparameter, $\nu$, for the Student's t-distribution likelihood distribution, as defined in section \ref{sec:student}.  This distribution peaks around 1.5, which means it is much closer to a Cauchy distribution than a Gaussian, so confirms the need for such a heavy-tailed distribution to adjust the fit to account for the outliers.
	
	\begin{figure}
		\begin{center}
			\includegraphics[width=0.8\textwidth]{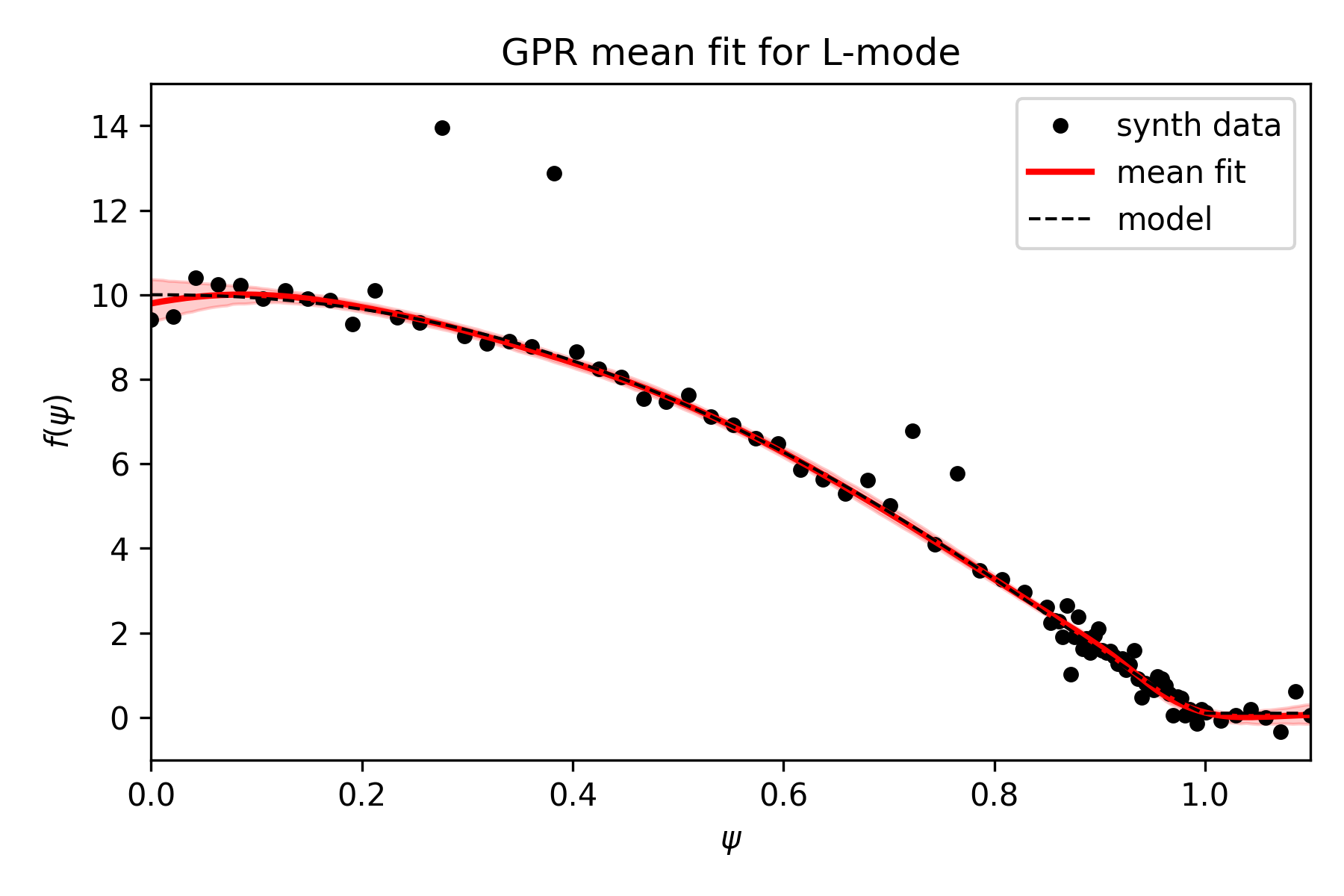}
		\end{center}
		\caption{Fit of an L-mode profile using full Bayesian approach with the change-point kernel described in section \ref{sec:kernel} and the Student's t-distribution likelihood described in \ref{sec:student}.}
		\label{fig:lmode-fit}
	\end{figure}
	
	\subsubsection{H-mode}
	
\begin{figure}
		\begin{center}
			\includegraphics[width=0.8\textwidth]{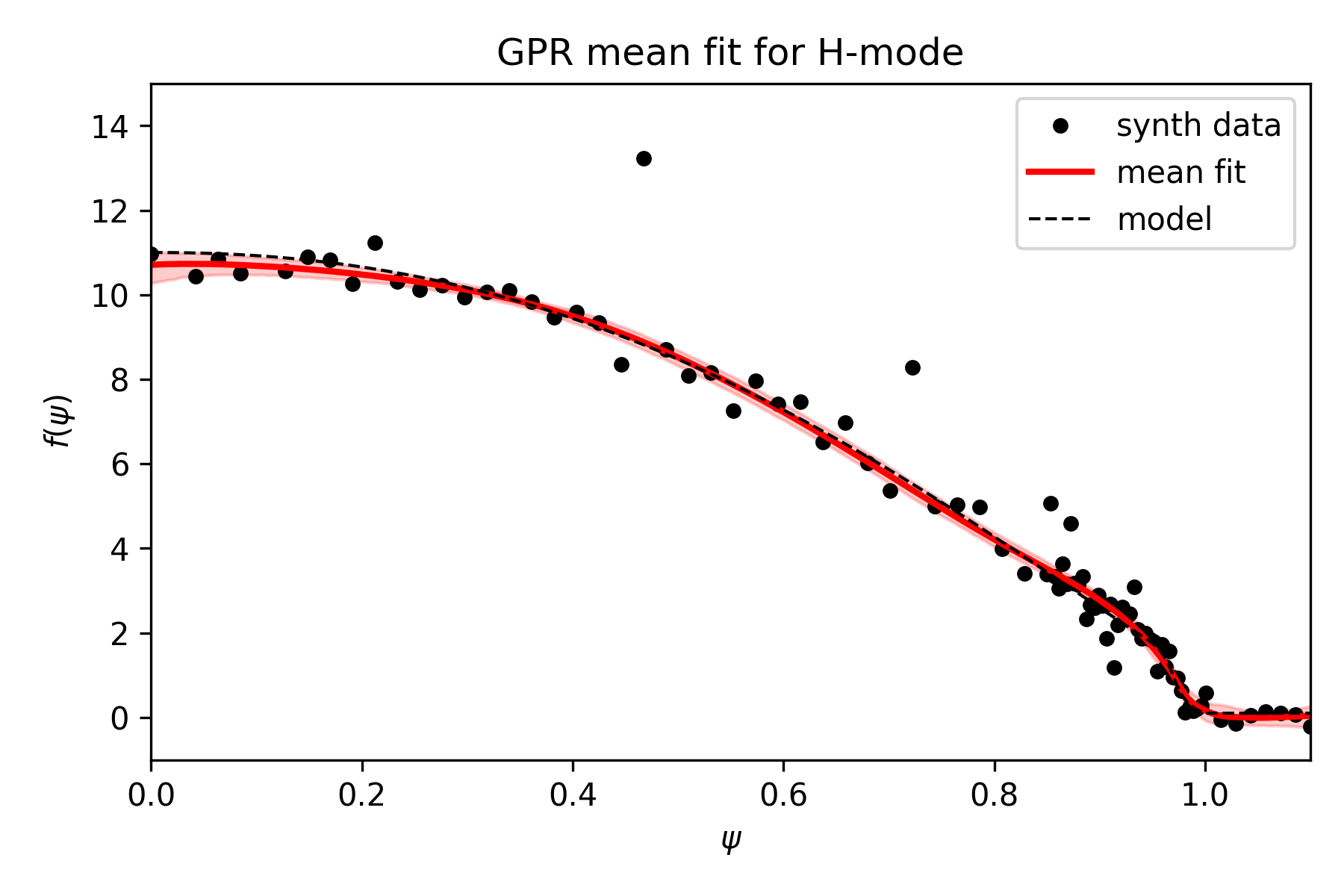}
		\end{center}
		\caption{Fit of an L-mode profile using full Bayesian approach with the change-point kernel described in section \ref{sec:kernel} and the Student's t-distribution likelihood described in \ref{sec:student}.}
		\label{fig:hmode-hp}
	\end{figure}

	The change in length scales of the pedestal region is the motivation for using the change point kernel discussed in Section~\ref{sec:kernel}.  Figure~\ref{fig:hmode-hp} shows the pedestal around $\psi=1$, and one can see that the remaining profile is simply the same as the L-mode profile with an offset.  The combination change point kernel allows this profile to be fit with the same kernel used to fit the L-mode profile in section \ref{sec:lmode}.
	
	\begin{figure}
		\begin{center}
			\includegraphics[width=\textwidth]{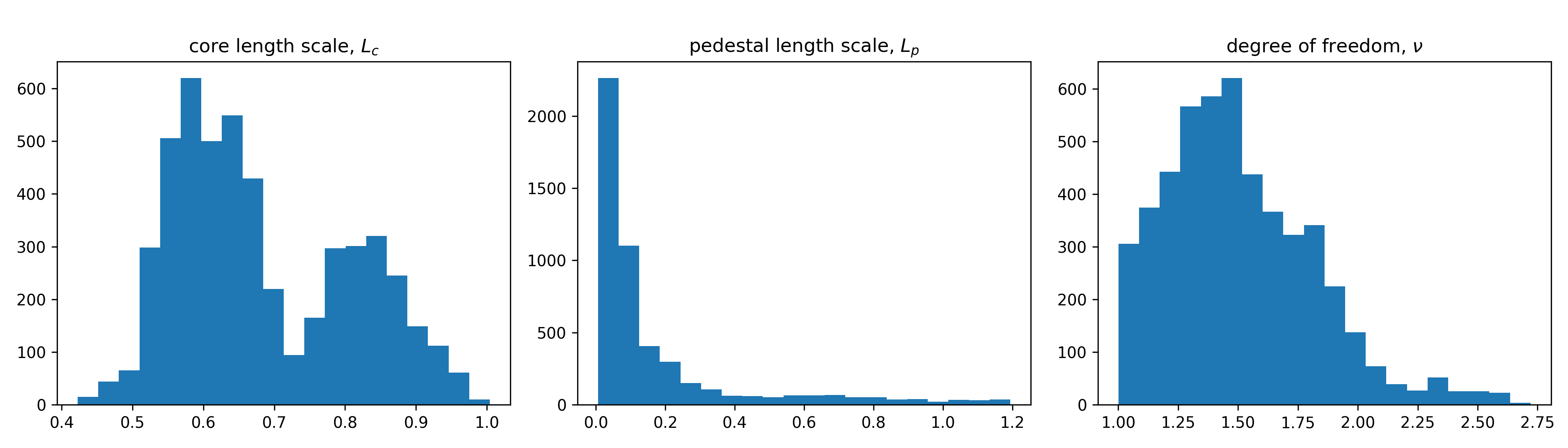}
		\end{center}
		\caption{Histograms of the hyperparameter distributions as found by the MCMC algorithm for the H-mode fit.  The y-axes are counts from the MCMC chain steps and the x-axes are the values for the variable in the title of the plots. Similar to the L-mode profile the noisy data enables several length scales to be determined in the core; however a single pedestal length scale is easily determined as the most probable.}
		\label{fig:hmode-histograms}
	\end{figure}
	
Figure~\ref{fig:hmode-histograms} shows the distributions for the hyperparameters for the H-mode fit.  The length scales distributions show the key feature of our kernel choice.  The hyperparameter $L_c$ is the hyperparameter for the correlation length in kernel A, which spans the majority of the spatial domain, and the majority of the values are between 0.5 and 1.  Interestingly, this is a bi-modal distribution confirming the need for the full Bayesian approach to sample this distribution.  In the $L_p$ distribution, which is the distribution for the correlation length hyperparameter in kernel B which spans $\psi \in [0.9,1]$, most of the values are much smaller, below 0.1.  This is what allows the fit to capture the gradient changes in the pedestal region equally as well as the rest of the domain is fit.  Again, like in the L-mode fit, the degree of freedom parameter, $\nu$, peaks around 1.5 so that outliers do not shift the mean fit.
	
	\subsubsection{H-mode with ITB}
\begin{figure}
		\begin{center}
			\includegraphics[width=0.8\textwidth]{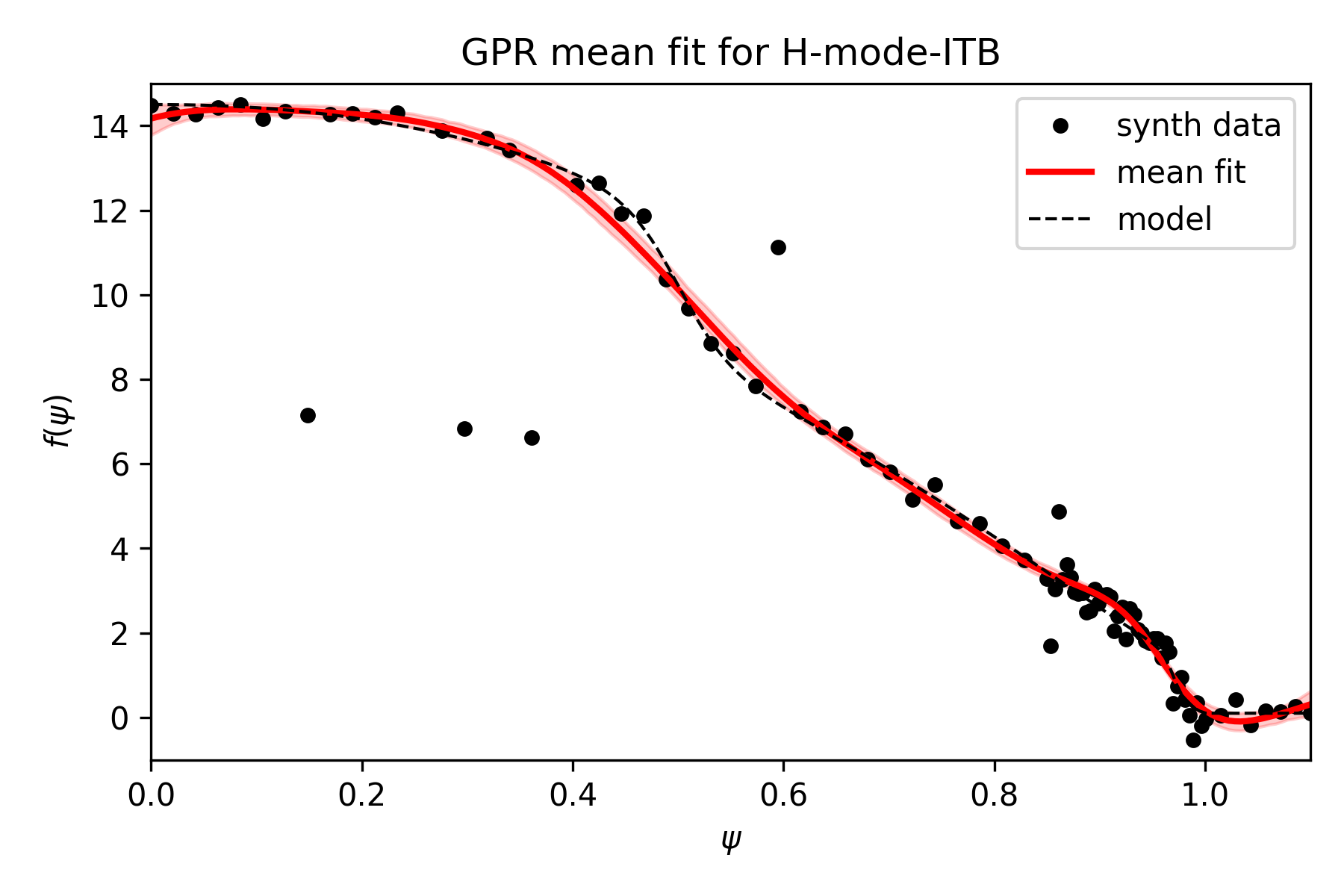}
		\end{center}
		\caption{Histograms of the hyperparameter distributions as found by the MCMC algorithm for the H-mode fit.  The y-axes are counts from the MCMC chain steps and the x-axes are the values for the variable in the title of the plots.  This illustrates the natural estimates of the fitting error inherent in using the Bayesian method.}
		\label{fig:hmode-itb-fit}
	\end{figure}

	The last case we look at explicitly is for an H-mode profile that contains an internal transport barrier (ITB).  This is another region of steep gradient but in the core of the plasma instead of the edge, where the pedestal can exist.  In our synthetic data, this will occur where the data is more sparse than the pedestal region, similar to real experiments.  This case is not explicitly fit by the kernel we introduce in section \ref{sec:kernel}, but it could be extended to do so by adding a third kernel that covers a region around $\psi=0.5$.  This would increase the number of hyperparameters, therefore the dimensionality of the parameter space through which the MCMC must sample.  It is still doable, but is saved for future work.  Since this is not done, the purpose of this section is to show that this method performs well even without these additions.  Figure \ref{fig:hmode-itb-fit} shows an example of a fit of this type of profile.  Though the gradient of the ITB is not perfectly fit by GPR, an inflection in the fit is seen as the fit attempts to capture the inflection in the underlying profile.  This fit better captures an ITB than a parameterized method for H-mode would, because there is no over-fitting, and yet requires no knowledge that such an ITB might exist.  This applies to any non-standard profile, as well - GPR is able to smoothly fit unexpected features in the underlying profile without making assumptions about tokamak regimes.
	
\subsection{Statistical analysis}
	
In this section we perform a statistical analysis of various GPR fitting methods on a large set of synthetic data that spans a broad parameter space in the profile shape and noise.  The four methods we test here cover two kernels, two likelihoods, and full vs empirical Bayes.  As described in section \ref{sec:intro}, empirical Bayes involves an analytic simplification assuming a Gaussian likelihood, allowing for simple hyperparameter optimization.  Full Bayesian, on the other hand, does not make this simplification so a method like Markov chain Monte Carlo must be used to sample the hyperparameter space (likelihood function is not restricted).  The Chilenski kernel \cite{chilenski2015} is used in the fusion community used for GPR regression of H-mode profiles for equilibrium reconstruction.  It has been implemented for use in OMFIT\cite{OMFIT2015} with OMFIT profiles~\cite{Logan2018}.  It is a standard Gibbs kernel, but with the length scale function defined and parameterized by a \emph{tanh} function to represent two length scales with a transition region.  We refer to this kernel and method as Chilenski.  We refer to the kernel proposed in section \ref{sec:kernel} as the change-point kernel in the following sections for brevity.  The four methods/cases we analyze are:
	\begin{enumerate}
	    \item Empirical Bayes with Chilenski kernel (Gibbs)
	    \item Empirical Bayes with change-point kernel (change-point)
	    \item Full Bayes with change-point kernel (implied Gaussian likelihood)
	    \item Full Bayes with change-point kernel and Student's t-distribution likelihood
	\end{enumerate}
It is possible for the Chilenski kernel to be used with empirical and full Bayesian methods, but herein we only use empirical Bayes with the Chilenski kernel.
	
	\subsubsection{Building database of fits}
	In order to analyze the quality of fit provided by the methods described in the previous sections, we have performed a large number of fits on a range of synthetic profiles.  There are two parts to doing a study like this.  First, a figure of merit must be defined to determine and compare the quality of the fits.  Secondly, a parameter space must be defined to generate the range of synthetic profiles.
	
	For the first, there are many figures of merit that can be used to describe the quality of fit.  We desire one that does not depend on the number of points in the resulting fit, so must therefore include a mean.  The root mean square error (RMSE) is a reasonable choice as it represents the standard deviation of the residuals between the fit and the true, underlying profile. It is important to note that the Chilenski method also aims to obtain high quality fit derivatives for transport model calculations. In this paper we are using the RMSE which only compares the quality of the data fit, not including any of the extra features involving derivatives that Chilenski developed. We define this as:
	
	\begin{equation}
	    RMSE = \sqrt{\frac{1}{N}\sum^N_i{(y_i-p_i)^2}}
	\end{equation}
	where $y_i$ is the value of the fit and $p_i$ is the true profile value.  Because these profiles are synthetically generated with noise added, we know the exact, underlying profile making this analysis straightforward and insightful beyond what we can know analyzing experimental data.
	
	Additionally, we have varied a series of parameter associated with the generation of synthetic data so that in the end we fit 5280 different profiles.  The set of parameters are
	\begin{enumerate}
	    \item noise amplitude: $\sigma_n \in [0.1, 0.15, 0.2, 0.25, 0.33]$
	    \item noise shift (systematic error): $y_{s} \in [0\%, 2\%, 5\%, 10\%]$
	    \item outlier number: $N_{OL} \in [0, 3, 5, 10]$
	    \item outlier amplitude: $\sigma_{OL} \in [2, 3, 4]$
	\end{enumerate}
	The number of points generated for these synthetic profiles is 88, so the number of outliers can also be written as a fraction of the total points: $N_{OL} \in [0\%, 2.2\%, 5.7\%, 11.3\%]$.  Profiles were generated for H-mode, L-mode, and H-mode with ITBs for these four parameter variations.  Additionally, for H-mode we added the following parameters
	\begin{enumerate}
	    \item edge density: $n_{edge} \in [0.01, 0.05, 0.1, 0.2]$ 
	    \item pedestal width: $w_{ped} \in [0.01, 0.015, 0.02]$
	\end{enumerate}
	
	Finally, for H-mode with ITB we added the following additional parameters
	\begin{enumerate}
	    \item ITB width: $w_{itb} \in [0.01, 0.015, 0.02]$
	    \item ITB height: $n_{itb} \in [0.5, 1.0, 1.5]$
	\end{enumerate}
	
	We then fit these synthetically generated profiles with four methods
	\begin{enumerate}
	    \item Chilenski kernel (Gibbs with \emph{tanh} length scale function) with empirical Bayes assumptions
	    \item Change-point  kernel (described in \ref{sec:kernel}) with empirical Bayes assumptions
	    \item Change-point kernel with Gaussian likelihood with full Bayesian approach 
	    \item Change-point kernel with Student's t-distribution likelihood with full Bayesian approach
	\end{enumerate}
	
	Between these profiles, methods, and the figure of merit we were able to build up a database of over 5000 sets of synethic data from that enable us to perform a quantitative comparison of the different methods.
	
	\subsubsection{Worst fits for each method}
	
	In order to put the RMSE values into context, we first want to show what constitutes a ``bad" fit using these methods.  An advantage of GPR is that when hyperparameter optimization is successful, even the bad fits are smooth and generally still acceptable fits.  This is not always the case for parameterized fitting methods that can fail spectacularly due to over-fitting.  To examine these bad cases, we have found the worst fit (highest RMSE) for each method and then plotted that fit along with the fit for the same data using all the other methods.  
	
\begin{figure} 
   \label{fig:worstfits}
		\begin{center}
			\includegraphics[width=\textwidth]{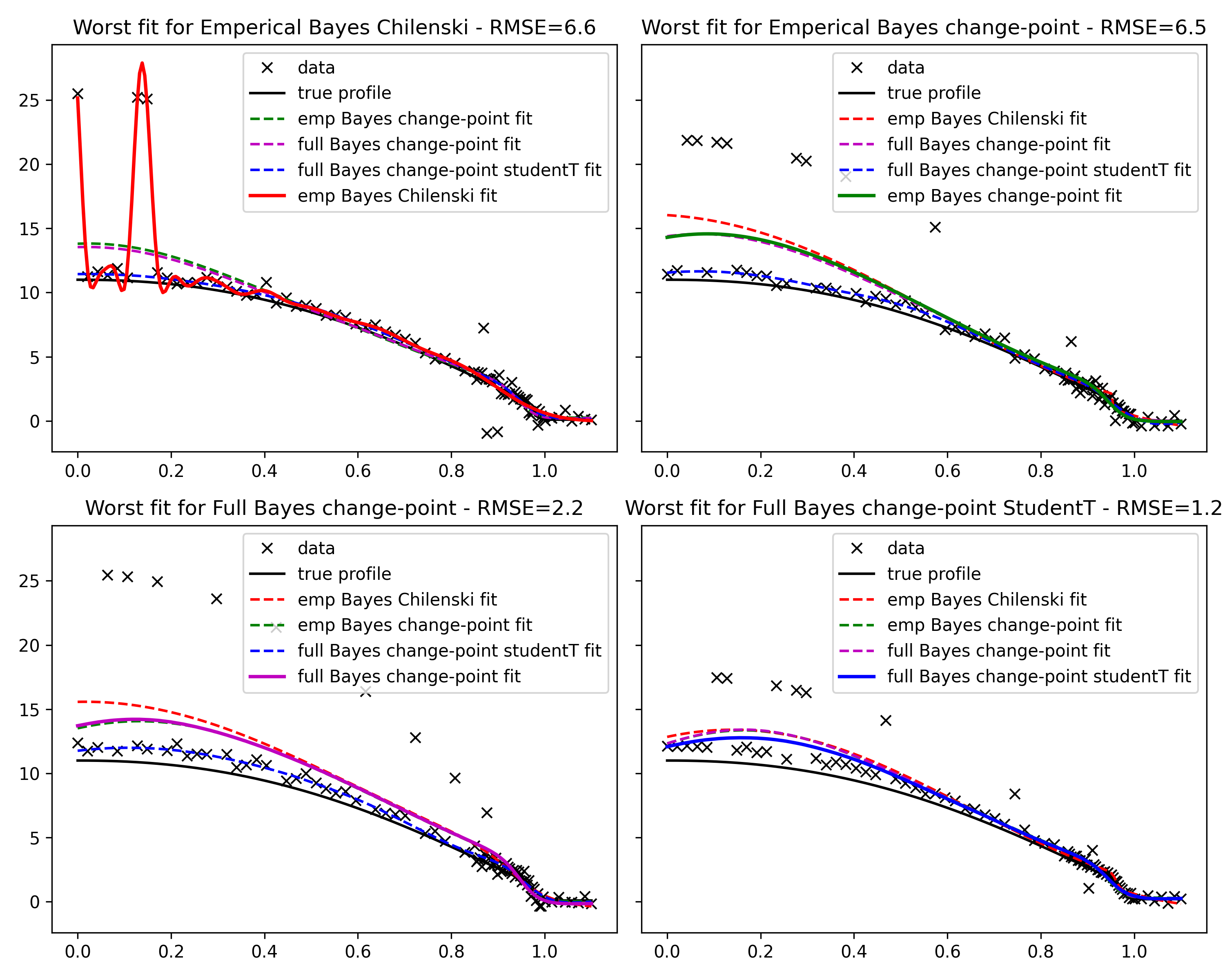}
		\end{center}
		\caption{Plots of the worst fit for each fitting method: Empirical Bayes Chilenski (top left), Empirical Bayes  (top right), Full Bayes (bottom left), and Full Bayes Student's t-distribution (bottom right) show the strengths and weakness of each method.  All methods struggle with the systematic error in the noise as represented by the shift of the noise.  The Full Bayesian method with the change-point kernel and Student's t distribution is the least sensitive to outliers and gives the most accurate fit however.  Note that in these cases, the outliers all lie above the profile in order to maintain positivity.}
\end{figure}
	
	These plots are shown in figure \ref{fig:worstfits}.  The worst fit using the empirical Bayesian method with the Chilenski kernel (top left plot) significantly over fit the data in the core and smoothed over the pedestal.  This is due to the outliers that are close to each other making the hyperparameter likelihood function multi-modal - we are most likely seeing a local maximum instead of the global.  This indicates a key problem that our outlier solution is able to address.  This is addressed by our method, but One could also imagine making the assumption that the core correlation length must always be larger than the pedestal/edge correlation length, thus constraining the Chilenski method so that this type of fit could not occur.  However, we invoked the Chilenski method using the same safeguards as in OMFIT \cite{OMFIT2015}, which would need to be modified for robustness.  The data in this fit are not particularly difficult to fit, as the other methods do a relatively good job in fitting them, though their means (excluding the Student's t-distribution likelihood method) are pulled above the true profile due to these outliers in the core region.  Therefore the strength of the Student's t-distribution likelihood is that it addressed multiple problems that outliers can cause in data fitting - affecting the mean and confusing the length scale hyperparameter optimization.
	
	The worst fit using the other three methods show outlying cases where the data happen to have all the outlying data points above the underlying profile.  This is an edge case since for the synthetic data the outliers have equal chances of lying above and below the true profile.  In all these cases the mean fit are pulled by the outliers away from the underlying profile by these outliers.  Additionally, these synthetic data profiles also sit in the parameter space where the noise offset is non-zero.  This is to be expected, since none of these methods are able to intrinsically detect systematic errors.
	
	Importantly, all of the proposed (non-Chelinski) methods, even in the worst case, capture the pedestal gradient well and therefore are able to identify pedestal location and width.  These are key parameters that EFIT uses to continue forward with the equilibrium calculation.  Additionally, even these worst cases result in smooth fits that would be unlikely to cause calculation errors as a part of an equilibrium reconstruction workflow.
	
	\subsubsection{Effect of outlier mitigation}
	
\begin{figure}
		\begin{center}
			\includegraphics[width=\textwidth]{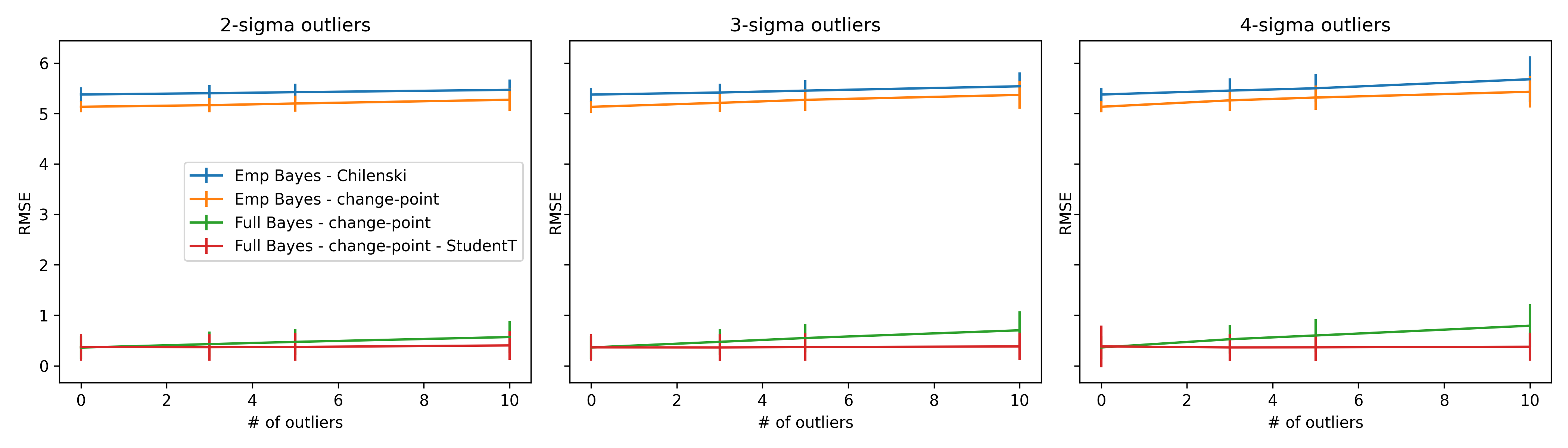}
		\end{center}
		\caption{Plots of the average and standard deviation for the RMSE of the fits across all synthetically generated profiles illustrate the effect of outliers on the RMSE for the fits.  Each fitting method is shown in a different color and the size of the outliers is $2\sigma$ (left), $3\sigma$ (middle) and $4\sigma$ (right).  Note the flatness of the fit using the Student's t-distribution likelihood, indicating a relative independence of the fit on number of outliers.}
		\label{fig:outliercomp}
	\end{figure}

	We introduce in section \ref{sec:student} a method for minimizing the impact of outliers on the fit by using a Student's t-distribution likelihood.  This heavy-tailed distribution accounts for outliers without shifting the mean.  Figure \ref{fig:outliercomp} shows the benefits of such an approach compared to the rest of the analyzed methods.
	
	It is clear from this plot that the full Bayesian approach provides much better fits on average than the empirical Bayesian approach.  Additionally, each of these lines is seen to trend up in error as more and more outliers are introduced, except for the method that includes outlier mitigation via the Student's t-distribution likelihood.  This method is roughly invariant to the number of outliers, at least in the range shown here which represents a maximum of over ~11\% outliers in the data.
	
	\subsubsection{Fitting L-mode versus H-mode}
	
	One of the objectives behind developing the fitting method described in this paper (change-point kernel with student-T likelihood) is to fit H-mode and L-mode equally well without having to know \emph{a priori} anything about the operational mode of the tokamak from which the data come.  This is accomplished through GPR with variable length scales in the kernel and hyperparameter optimization/sampling.  However, we want to test how well these methods perform at fitting L-mode and H-mode to make sure that the methods are as effective as they should be.
	
\begin{figure}
		\begin{center}
			\includegraphics[width=\textwidth]{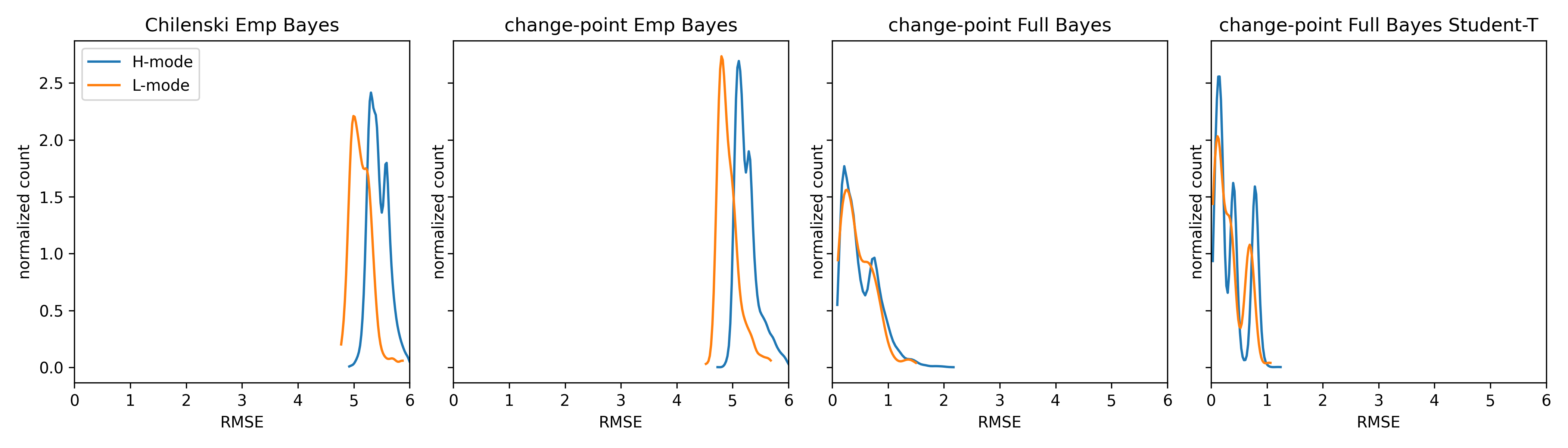}
		\end{center}
		\caption{These plots show the distribution of RMSE for the various methods for H-mode and L-mode.  It is important to note that the RMSE axes differ in scale for the plots - the full Bayesian approach consistently produces fits with lower RMSE than the empirical Bayesian approach.}
		\label{fig:modeComp}
	\end{figure}

	Figure~\ref{fig:modeComp} shows the distribution of RMSE for both L-mode and H-mode separately, and for each fitting method separately.  There are a few key points worth emphasizing from these plots.  Firstly, it's clear that the empirical Bayesian does a worse job, regardless of kernel, for fitting H-mode than it does for fitting L-mode.  This is likely due to the large parameter space involved and ending up in a local maximum instead of global for the likelihood maximization.  Also, the fit for H-mode is more difficult as the gradients are steeper, so the parameter space is likely also steeper and multi-modal.  Secondly, the Full Bayesian approach sees minimal different between fitting L-mode and H-mode, and the overall RMSE is lower for these than for the empirical Bayesian fits.  The multi-modal behavior seen in this figure is due to the large parameter space encapsulated within the data.  The peaks at higher RMSE are due to noise offsets and large outliers, while the peaks at smaller RMSE are easier profiles to fit.
	
\section{Discussion}
	
This paper used synthetic data with the ability to control for the the amount of noise, the degree of outliers, and the effect of systematic bias to perform a statistical analysis of various GPR methods.  These methods include the non-stationary empirical Bayes method introduced by Chilenski~\cite{chilenski2015}, as well as an alternative to non-stationary hyperparameters:  the change-point kernel (a linear combination of two stationary kernels).  The effect is similar, however, in that it allows multiple length scales to be chosen for GPR and applied to different locations on the $\psi$ axis.  The change-point kernel we introduced is, however, more flexible than a Gibbs kernel with a \emph{tanh} length scale function as it allows any number of length scales that can change at arbitrary positions in $\psi$.  
	
The most important new method introduced here is the Student's t-distribution for the likelihood function.  This allows the fitting to handle outliers reliably and without any prior knowledge as to which points are outliers and how many there may be.  The use of the Student's t-distribution rules out the Empirical Bayes method used here, but is easily implemented in our Full Bayesian method using GPflow.  We compared Full Bayesian with the standard Gaussian likelihood to the Student's t-distribution and found that both were more accurate than the Empirical Bayesian techniques.
	
To be truly quantitative in our judgements, a statistical analysis using the various methods we described to compare their effectiveness across a range of profiles including L-mode, H-mode, and H-mode with ITB.  This also included a variety of settings on the noise including amplitude, mean, outlier number, etc.  to allow an analysis of over 5,000 sets of generated data.   However, fundamentally the synthetic data is generated with Gaussian noise such that it should match well to the commonly used square exponential kernel.   The noise was varied to allow a shifted Gaussian, representing systematic errors in a diagnostic, as well as outliers, representing bad channels in a diagnostic.  

The most important results of this statistical study can be seen in Figure~\ref{fig:outliercomp}.   H-mode profiles are consistently more difficult to difficult to fit with empirical Bayes method, with a consistent RMS error that is approximately 10\% higher than L-mode.  There are also differences between L-mode and H-mode profiles in the Full Bayesian methods, but the differences are slight.  The change-point kernel is slightly better than the non-stationary hyperparameter \emph{tanh} profiles of Chilenski, but the difference is slight.  Full Bayesian methods are an order of magnitude more accurate Empirical Bayesian methods, but comes at a computational cost that is roughly an order of magnitude more expensive.  Though we did not test it, we would expect the Chilenski kernel to perform similarly to our own using the full Bayesian method since it also performed similar to our own using empirical Bayes.  The Student's t-distribution function for the likelihood is roughly twice as accurate as the Gaussian likelihood when there are a large number of outliers.  The lack of sensitivity to outliers by the Student's t-distribution can visually be seen in Figure~\ref{fig:worstfits}. It is important to note that  even the worst fits provided by our proposed methods result in smooth, reasonable fits to the data and are robust.
	
In future work, we will continue to refine these methods as we apply them to experimental data in a similar statistical manor.  When working with experimental data, however, we do not have the luxury of knowing the nature of the true profile.  Fortunately, the study that we have done here can give us confidence that we are able to capture the underlying profile with a fit, barring the existence of systematic offset in the noise.
	
	\subsubsection*{Acknowledgements}
    This material is based upon work supported by the Department of Energy under Award Number(s) DE-SC0021380.  We wish to thank Drs. Lang Lao, Joseph McClenaghan, and the EFIT-AI Team for useful discussions related to this work.
	
	\bibliographystyle{unsrt} 
	\bibliography{master}

\begin{thebibliography}{10}

\bibitem{HazeltineMeiss}
R.D. Hazeltine and J.D. Meiss.
\newblock {\em Plasma Confinement}.
\newblock Addison-Wesley Publishing Company, Redwood City, CA, 1992.

\bibitem{lao85}
L~L Lao, H~St John, R~D Stambaugh, A~G Kellman, and W~Pfeiffer.
\newblock {Reconstruction of current profile parameters and plasma shapes in
  tokamaks}.
\newblock {\em Nuclear Fusion}, 25(11):1611, 1985.

\bibitem{lao05}
LL~Lao, HE~St John, Q~Peng, JR~Ferron, EJ~Strait, TS~Taylor, WH~Meyer, C~Zhang,
  and KI~You.
\newblock Mhd equilibrium reconstruction in the diii-d tokamak.
\newblock {\em Fusion science and technology}, 48(2):968--977, 2005.

\bibitem{ferron1998real}
JR~Ferron, ML~Walker, LL~Lao, HE~St John, DA~Humphreys, and JA~Leuer.
\newblock Real time equilibrium reconstruction for tokamak discharge control.
\newblock {\em Nuclear fusion}, 38(7):1055, 1998.

\bibitem{Strait_1994}
E.~J. Strait.
\newblock Stability of high beta tokamak plasmas*.
\newblock {\em Physics of Plasmas}, 1(5):1415--1431, 1994.

\bibitem{Levinton_1995}
F.~M. Levinton, M.~C. Zarnstorff, S.~H. Batha, M.~Bell, R.~E. Bell, R.~V.
  Budny, C.~Bush, Z.~Chang, E.~Fredrickson, A.~Janos, J.~Manickam, A.~Ramsey,
  S.~A. Sabbagh, G.~L. Schmidt, E.~J. Synakowski, and G.~Taylor.
\newblock Improved confinement with reversed magnetic shear in tftr.
\newblock {\em Phys. Rev. Lett.}, 75:4417--4420, Dec 1995.

\bibitem{Strait_1995}
E.~J. Strait, L.~L. Lao, M.~E. Mauel, B.~W. Rice, T.~S. Taylor, K.~H. Burrell,
  M.~S. Chu, E.~A. Lazarus, T.~H. Osborne, S.~J. Thompson, and A.~D. Turnbull.
\newblock Enhanced confinement and stability in diii-d discharges with reversed
  magnetic shear.
\newblock {\em Phys. Rev. Lett.}, 75:4421--4424, Dec 1995.

\bibitem{Zwingmann_2003}
W~Zwingmann.
\newblock Equilibrium analysis of steady state tokamak discharges.
\newblock {\em Nuclear Fusion}, 43(9):842--850, aug 2003.

\bibitem{Lao_1985}
L.L. Lao, H.~St. John, R.D. Stambaugh, A.G. Kellman, and W.~Pfeiffer.
\newblock Reconstruction of current profile parameters and plasma shapes in
  tokamaks.
\newblock {\em Nuclear Fusion}, 25(11):1611--1622, nov 1985.

\bibitem{Lao_2005}
L.~L. Lao, H.~E.~St. John, Q.~Peng, J.~R. Ferron, E.~J. Strait, T.~S. Taylor,
  W.~H. Meyer, C.~Zhang, and K.~I. You.
\newblock Mhd equilibrium reconstruction in the diii-d tokamak.
\newblock {\em Fusion Science and Technology}, 48(2):968--977, 2005.

\bibitem{Lao_1990}
L.L. Lao, J.R. Ferron, R.J. Groebner, W.~Howl, H.~St. John, E.J. Strait, and
  T.S. Taylor.
\newblock Equilibrium analysis of current profiles in tokamaks.
\newblock {\em Nuclear Fusion}, 30(6):1035--1049, jun 1990.

\bibitem{In_2000}
Y~In, J.J Ramos, A.E Hubbard, I.H Hutchinson, M~Porkolab, J~Snipes, S.M Wolfe,
  and A~Bondeson.
\newblock Resistiven= 1 modes in reversed magnetic shear alcator c-mod plasmas.
\newblock {\em Nuclear Fusion}, 40(8):1463--1468, aug 2000.

\bibitem{OBrien_1992}
D.P O{\textquotesingle}Brien, L.L Lao, E.R Solano, M~Garribba, T.S Taylor, J.G
  Cordey, and J.J Ellis.
\newblock Equilibrium analysis of iron core tokamaks using a full domain
  method.
\newblock {\em Nuclear Fusion}, 32(8):1351--1360, aug 1992.

\bibitem{Nelson_1994}
B.~A. Nelson, T.~R. Jarboe, D.~J. Orvis, L.~A. McCullough, J.~Xie, C.~Zhang,
  and L.~Zhou.
\newblock Formation and sustainment of a 150 ka tokamak by coaxial helicity
  injection.
\newblock {\em Phys. Rev. Lett.}, 72:3666--3669, Jun 1994.

\bibitem{Sykes_1999}
A~Sykes, the START~Team, the NBI~Team, the MAST~Team, and the Theory~Team.
\newblock The spherical tokamak programme at culham.
\newblock {\em Nuclear Fusion}, 39(9Y):1271--1281, sep 1999.

\bibitem{Appel_2001}
L.C Appel, M.K Bevir, and M.J Walsh.
\newblock Equilibrium reconstruction in the {START} tokamak.
\newblock {\em Nuclear Fusion}, 41(2):169--180, feb 2001.

\bibitem{Appel_2006}
LC~Appel, GTA Huysmans, LL~Lao, PJ~McCarthy, DG~Muir, ER~Solano, J~Storrs,
  D~Taylor, W~Zwingmann, et~al.
\newblock A unified approach to equilibrium reconstruction.
\newblock In {\em Proceedings-33rd EPS conference on Controlled Fusion and
  Plasma Physics, pp. P--2.160}, 2006.

\bibitem{Lee_1999}
Bongju Lee, Neil Pomphrey, and Lang~L. Lao.
\newblock Physics design of poloidal field, toroidal field, and external
  magnetic diagnostics in kstar.
\newblock {\em Fusion Technology}, 36(3):278--288, 1999.

\bibitem{Park_2011}
Y.S. Park, S.A. Sabbagh, J.W. Berkery, J.M. Bialek, Y.M. Jeon, S.H. Hahn,
  N.~Eidietis, T.E. Evans, S.W. Yoon, J.-W. Ahn, J.~Kim, H.L. Yang, K.-I. You,
  Y.S. Bae, J.~Chung, M.~Kwon, Y.K. Oh, W.-C. Kim, J.Y. Kim, S.G. Lee, H.K.
  Park, H.~Reimerdes, J.~Leuer, and M.~Walker.
\newblock {KSTAR} equilibrium operating space and projected stabilization at
  high normalized beta.
\newblock {\em Nuclear Fusion}, 51(5):053001, apr 2011.

\bibitem{Oikawa_2000}
T~Oikawa, K~Ushigusa, C.B Forest, M~Nemoto, O~Naito, Y~Kusama, Y~Kamada,
  K~Tobita, S~Suzuki, T~Fujita, H~Shirai, T~Fukuda, M~Kuriyama, T~Itoh,
  Y~Okumura, K~Watanabe, L~Grisham, and JT-60 Team.
\newblock Heating and non-inductive current drive by negative ion based {NBI}
  in {JT}-60u.
\newblock {\em Nuclear Fusion}, 40(3Y):435--443, mar 2000.

\bibitem{Sabbagh_2001}
S.A Sabbagh, S.M Kaye, J~Menard, F~Paoletti, M~Bell, R.E Bell, J.M Bialek,
  M~Bitter, E.D Fredrickson, D.A Gates, A.H Glasser, H~Kugel, L.L Lao, B.P
  LeBlanc, R~Maingi, R.J Maqueda, E~Mazzucato, D~Mueller, M~Ono, S.F Paul,
  M~Peng, C.H Skinner, D~Stutman, G.A Wurden, W~Zhu, and NSTX~Research Team.
\newblock Equilibrium properties of spherical torus plasmas in {NSTX}.
\newblock {\em Nuclear Fusion}, 41(11):1601--1611, nov 2001.

\bibitem{Li_2011}
Y.~G. Li, Ph. Lotte, W.~Zwingmann, C.~Gil, and F.~Imbeaux.
\newblock Efit equilibrium reconstruction including polarimetry measurements on
  tore supra.
\newblock {\em Fusion Science and Technology}, 59(2):397--405, 2011.

\bibitem{Li_2003}
J.~Li, B.~N. Wan, J.~R. Luo, G.~L. Kuang, Y.~P. Zhao, J.~Y. Zhao, X.~D. Zhang,
  X.~N. Liu, P.~Fu, J.~K. Xie, C.~Zhang, X.~M. Gu, J.~S. Mao, J.~F. Shan, H.~Y.
  Bai, K.~Gentle, B.~Rowan, P.~Philippe, H.~Huang, L.~Lao, V.~Chan, T.~Watari,
  T.~Seki, and N.~Nakamura.
\newblock Long pulse enhanced confinement discharges in the ht-7
  superconducting tokamak by ion bernstein wave heating and lower hybrid wave
  current drive.
\newblock {\em Physics of Plasmas}, 10(5):1653--1658, 2003.

\bibitem{Hongda_2006}
He~Hongda, Zhang Jinhua, Dong Jiaqi, and Li~Qiang.
\newblock Study of plasma {MHD} equilibrium in {HL}-2a tokamak.
\newblock {\em Plasma Science and Technology}, 8(4):397--401, jul 2006.

\bibitem{Jinping_2009}
Qian Jinping, Wan Baonian, L.~L Lao, Shen Biao, S.~A Sabbagh, Sun Youwen, Liu
  Dongmei, Xiao Bingjia, Ren Qilong, Gong Xianzu, and Li~Jiangang.
\newblock Equilibrium reconstruction in {EAST} tokamak.
\newblock {\em Plasma Science and Technology}, 11(2):142--145, apr 2009.

\bibitem{Li_2013}
G~Q Li, Q~L Ren, J~P Qian, L~L Lao, S~Y Ding, Y~J Chen, Z~X Liu, B~Lu, and
  Q~Zang.
\newblock Kinetic equilibrium reconstruction on {EAST} tokamak.
\newblock {\em Plasma Physics and Controlled Fusion}, 55(12):125008, nov 2013.

\bibitem{Carlstrom92}
T.~N. Carlstrom, G.~L. Campbell, J.~C. DeBoo, R.~Evanko, J.~Evans, C.~M.
  Greenfield, J.~Haskovec, C.~L. Hsieh, E.~McKee, R.~T. Snider, R.~Stockdale,
  P.~K. Trost, and M.~P. Thomas.
\newblock Design and operation of the multipulse thomson scattering diagnostic
  on diii‐d (invited).
\newblock {\em Rev. Sci. Inst.}, 63(10):4901--4906, 1992.

\bibitem{Carlstrom_1995}
T.~N. Carlstrom, J.~H. Foote, D.~G. Nilson, and B.~W. Rice.
\newblock Design of the divertor thomson scattering system on diii‐d.
\newblock {\em Review of Scientific Instruments}, 66(1):493--495, 1995.

\bibitem{Christiansen_1989}
JP~Christiansen, JD~Callen, JJ~Ellis, and RS~Granetz.
\newblock Determination of current distribution in jet from soft x-ray
  measurements.
\newblock {\em Nuclear fusion}, 29(5):703, 1989.

\bibitem{fischer2004}
R.~Fischer and A.~Dinklage.
\newblock Integrated data analysis of fusion diagnostics by means of the
  bayesian probability theory.
\newblock {\em Review of Scientific Instruments}, 75(10):4237--4239, October
  2004.

\bibitem{fischer2004b}
R.~Fischer.
\newblock Bayesian experimental design --- studies for fusion diagnostics.
\newblock In {\em AIP Conference Proceedings}. AIP, 2004.

\bibitem{williams2006gaussian}
Christopher~K Williams and Carl~Edward Rasmussen.
\newblock {\em Gaussian processes for machine learning}.
\newblock MIT press Cambridge, MA, 2006.

\bibitem{svensson2011non}
Jakob Svensson.
\newblock {\em Non-parametric tomography using Gaussian processes}.
\newblock EFDA, 2011.

\bibitem{li2016}
Dong Li, Y.~B. Dong, Wei Deng, Z.~B. Shi, B.~Z. Fu, J.~M. Gao, T.~B. Wang, Yan
  Zhou, Yi~Liu, Q.~W. Yang, and X.~R. Duan.
\newblock Bayesian tomography and integrated data analysis in fusion
  diagnostics.
\newblock {\em Rev. Sci. Instrum.}, 87(11):11E319, August 2016.

\bibitem{joung2018}
Semin Joung, Jaewook Kim, Sehyun Kwak, Kyeo reh Park, S.~H. Hahn, H.~S. Han,
  H.~S. Kim, J.~G. Bak, S.~G. Lee, and Y.~c.~Ghim.
\newblock Imputation of faulty magnetic sensors with coupled bayesian and
  gaussian processes to reconstruct the magnetic equilibrium in real time.
\newblock {\em Review of Scientific Instruments}, 89(10):10K106, oct 2018.

\bibitem{kwak2020bayesian}
Sehyun Kwak, Jakob Svensson, S~Bozhenkov, Joanne Flanagan, Mark Kempenaars,
  Alexandru Boboc, Y-C Ghim, and JET Contributors.
\newblock Bayesian modelling of thomson scattering and multichannel
  interferometer diagnostics using gaussian processes.
\newblock {\em Nuclear Fusion}, 60(4):046009, 2020.

\bibitem{kwak2021bayesian}
Sehyun Kwak, J.~Svensson, S.~Bozhenkov, H.~Trimino Mora, U.~Hoefel, A.~Pavone,
  M.~Krychowiak, A.~Langenberg, Y.~c.~Ghim, and W7-X Team.
\newblock Bayesian modelling of multiple plasma diagnostics at wendelstein 7-x.
\newblock Mar 2021.

\bibitem{chilenski2015}
M.A. Chilenski, M.~Greenwald, Y.~Marzouk, N.T. Howard, A.E. White, J.E. Rice,
  and J.R. Walk.
\newblock Improved profile fitting and quantification of uncertainty in
  experimental measurements of impurity transport coefficients using gaussian
  process regression.
\newblock {\em Nucl. Fusion}, 55(2):023012, January 2015.

\bibitem{fischer2020}
R.~Fischer, L.~Giannone, J.~Illerhaus, P.~J. McCarthy, R.~M. McDermott, and
  ASDEX~Upgrade Team.
\newblock Estimation and uncertainties of profiles and equilibria for fusion
  modeling codes.
\newblock {\em Fusion Science and Technology}, 76(8):879--893, nov 2020.

\bibitem{ho2019}
Aaron Ho, Jonathan Citrin, Fulvio Auriemma, Clarisse Bourdelle, Francis~J
  Casson, Hyun-Tae Kim, Pierre Manas, Gabor Szepesi, Henri Weisen, and JET
  Contributors.
\newblock Application of gaussian process regression to plasma turbulent
  transport model validation via integrated modelling.
\newblock {\em Nuclear fusion}, 59(5):056007, 2019.

\bibitem{mathews2021}
Abhilash Mathews and Jerry~W Hughes.
\newblock Quantifying experimental edge plasma evolution via multidimensional
  adaptive gaussian process regression.
\newblock {\em IEEE Transactions on Plasma Science}, 49(12):3841--3847, 2021.

\bibitem{Howell_2020}
Eric~C Howell and JD~Hanson.
\newblock Development of a non-parametric gaussian process model in the
  three-dimensional equilibrium reconstruction code v3fit.
\newblock {\em Journal of Plasma Physics}, 86(1), 2020.

\bibitem{kwak2021equilibria}
Sehyun Kwak, J.~Svensson, O.~Ford, L.~Appel, Y.~c.~Ghim, and JET Contributors.
\newblock Bayesian equilibria of axisymmetric plasmas.
\newblock Mar 2021.

\bibitem{sciortino2020}
F~Sciortino, NT~Howard, ES~Marmar, T~Odstrcil, NM~Cao, R~Dux, AE~Hubbard,
  JW~Hughes, JH~Irby, YM~Marzouk, et~al.
\newblock Inference of experimental radial impurity transport on alcator c-mod:
  Bayesian parameter estimation and model selection.
\newblock {\em Nuclear Fusion}, 60(12):126014, 2020.

\bibitem{xing2021}
Z.A. Xing, D.~Eldon, A.O. Nelson, M.A. Roelofs, W.J. Eggert, O.~Izacard, A.S.
  Glasser, N.C. Logan, O.~Meneghini, S.P. Smith, R.~Nazikian, and E.~Kolemen.
\newblock {CAKE}: Consistent automatic kinetic equilibrium reconstruction.
\newblock {\em Fusion Engineering and Design}, 163:112163, feb 2021.

\bibitem{OMFIT2015}
O.~Meneghini, S.P. Smith, L.L. Lao, O.~Izacard, Q.~Ren, J.M. Park, J.~Candy,
  Z.~Wang, C.J. Luna, V.A. Izzo, B.A. Grierson, P.B. Snyder, C.~Holland,
  J.~Penna, G.~Lu, P.~Raum, A.~McCubbin, D.M. Orlov, E.A. Belli, N.M. Ferraro,
  R.~Prater, T.H. Osborne, A.D. Turnbull, and G.M. Staebler.
\newblock Integrated modeling applications for tokamak experiments with omfit.
\newblock {\em Nuclear Fusion}, 55(8):083008, 2015.

\bibitem{Logan2018}
N.~C. Logan, B.~A. Grierson, S.~R. Haskey, S.~P. Smith, O.~Meneghini, and
  D.~Eldon.
\newblock Omfit tokamak profile data fitting and physics analysis.
\newblock {\em Fusion Science and Technology}, 74(1-2):125--134, 2018.

\bibitem{Note1}
This paper discusses the OMFIT profiles module that can also call out to the
  \protect \texttt {gptools} software package that forms the basis of the paper
  by Chilenski~\cite {chilenski2015}.

\bibitem{Gibbs1997}
M.N. Gibbs.
\newblock Bayesian gaussian processes for regression and classification.
\newblock {\em PhD thesis}, 1997.

\bibitem{buckingham2008comparative}
Lawrence Buckingham, James~M Hogan, Paul Roe, Jiro Sumitomo, and Michael
  Towsey.
\newblock Comparative studies made simple in gpflow.
\newblock In {\em 2008 Eighth IEEE International Symposium on Cluster Computing
  and the Grid (CCGRID)}, pages 699--699. IEEE, 2008.

\end{thebibliography}
	
\end{document}